\documentclass[a4paper,fleqn]{cas-dc}

\usepackage[numbers]{natbib}
\usepackage[utf8]{inputenc}
\DeclareUnicodeCharacter{22EF}{\dots}

\def\tsc#1{\csdef{#1}{\textsc{\lowercase{#1}}\xspace}}
\tsc{WGM}
\tsc{QE}
\tsc{EP}
\tsc{PMS}
\tsc{BEC}
\tsc{DE}

\begin{document}
\let\WriteBookmarks\relax
\def\floatpagepagefraction{1}
\def\textpagefraction{.001}
\shorttitle{OLi$_3$-decorated Irida-graphene}
\shortauthors{Laranjeira et~al.}

\title [mode = title]{OLi$_3$-decorated Irida-graphene for High-capacity Hydrogen Storage: A First-principles Study}

\author[1]{José A. S. Laranjeira}
\affiliation[1]{organization={Modeling and Molecular Simulation Group},
    addressline={São Paulo State University (UNESP), School of Sciences}, 
    city={Bauru},
    postcode={17033-360}, 
    state={SP},
    country={Brazil}}

\credit{Data curation, Formal analysis, Writing -- review \& editing, Writing -- original draft}

\author[2]{Warda Elaggoune}
\credit{Data curation, Formal analysis, Writing -- review \& editing, Writing -- original draft}
\affiliation[2]{organization={Laboratoire de Physique des Matériaux (L2PM), Faculté des mathématiques, de l'informatique et des sciences de la matière, Université 8 Mai 1945, BP 401, Guelma, Algeria}} 

\cormark[1] 
\cortext[1]{Corresponding author: elaggoune.warda@univ-guelma.dz}

\author[1]{Nicolas F. Martins}
\credit{Investigation, Formal analysis, Writing -- original draft, Writing -- review \& editing}

\author[3]{Xihao Chen}
\affiliation[3]{organization={School of Materials Science and Engineering},
    addressline={Chongqing University of Arts and Sciences}, 
    city={Chongqing},
    postcode={402160}, 
    country={China}}
\credit{Investigation, Formal analysis, Writing -- review \& editing}

\author[1]{Julio R. Sambrano}
\cormark[2] 
\cortext[2]{Main corresponding author: jr.sambrano@unesp.br}
\credit{Data curation, Formal analysis, Writing -- review \& editing}

\begin{abstract}
Efficient hydrogen storage in solid-state materials is essential for next-generation energy systems, yet achieving a high gravimetric capacity with optimal adsorption characteristics remains a critical challenge. Although Li-decorated irida-graphene (IG) has shown promising hydrogen storage potential, its capacity is limited to $\sim$ 7wt\%, which, despite exceeding the U.S. DOE target, remains inadequate for large-scale applications. Additionally, Li clustering over extended cycles may compromise adsorption efficiency and structural stability. In this study, we employ first-principles calculations to investigate the hydrogen storage potential of IG decorated with superalkali OLi$_3$ clusters, aiming to enhance the adsorption capacity and stability for advanced hydrogen storage technologies. Our findings show that the OLi$_3$ clusters exhibit a significant binding energy of -3.24 eV, which highlights its strong interaction with the IG. OLi$_3$@IG complex can host up to 12H$_2$ molecules, with optimal maximum storage capacity of 10.00 wt\%. Additionally, the release temperature (T$_R$) and \textit{ab initio} molecular dynamics (AIMD) simulations indicate that H$_2$ molecules can be efficiently released at operating temperatures under ambient conditions. These results highlight the potential of OLi$_3$-decorated irida-graphene as a promising candidate for reversible hydrogen storage. 

\end{abstract}


\begin{graphicalabstract}
\begin{figure}[h!]
    \centering
    \includegraphics [width=1\textwidth] {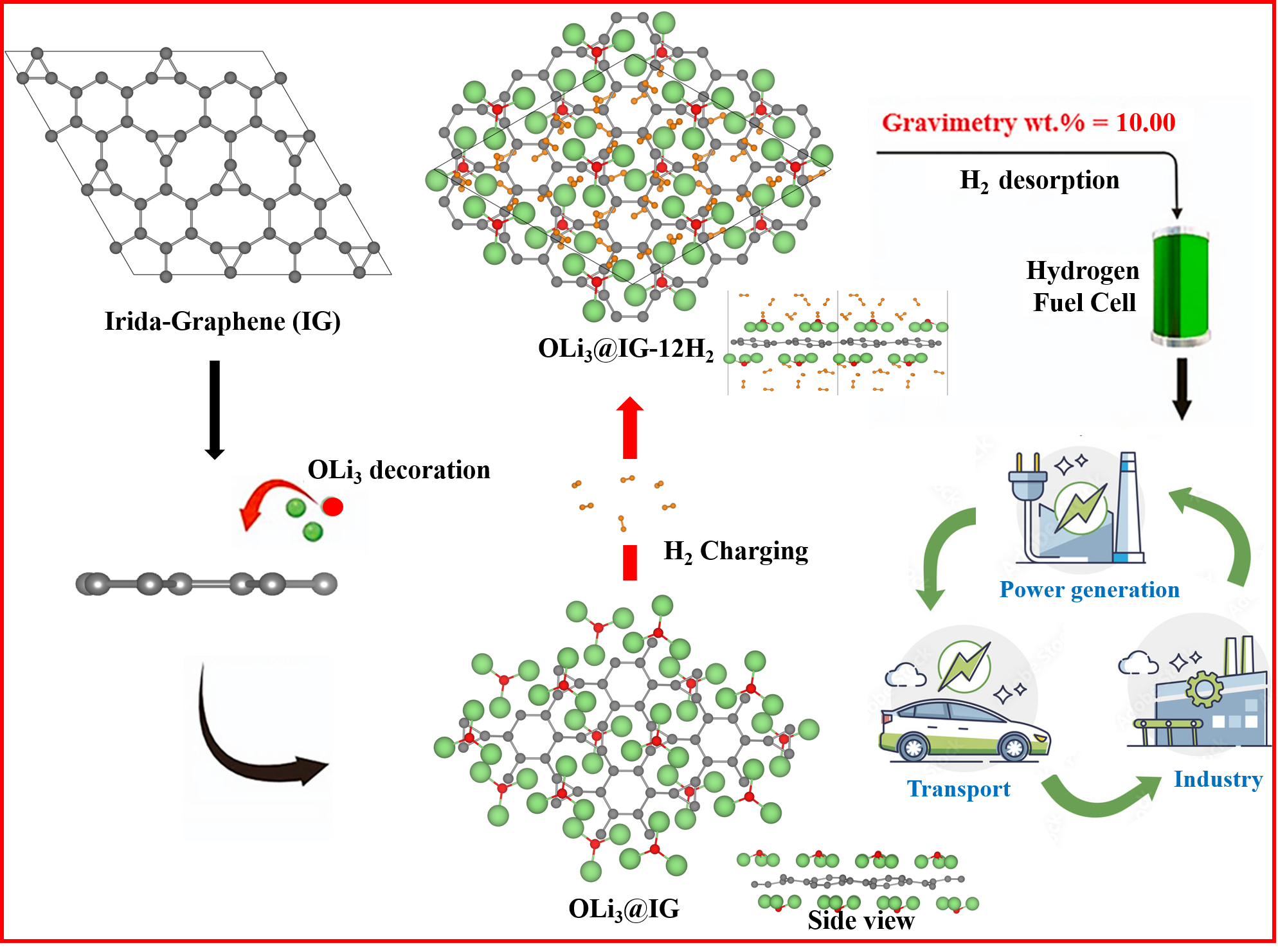} 
\end{figure}

\end{graphicalabstract}

\begin{highlights}
\item The H$_2$ adsorption behavior of OLi$_3$-decorated irida-graphene (IG) was investigated using the DFT method.
\item OLi$_3$-decorated IG could adsorb 12H$_2$ molecules, achieving 10.00 wt\% storage, exceeding the U.S. DOE target.
\item AIMD simulations and release temperature analysis support the reversible nature of hydrogen storage.
\end{highlights}


\maketitle

\section{Introduction}

The global energy crisis, coupled with environmental concerns such as climate change and greenhouse gas emissions, has intensified the demand for sustainable and emission-free energy alternatives \cite{lenferna2018can}. The dependence of fossil fuels has led to severe environmental degradation and resource depletion, which requires urgent development of clean energy solutions. Among the various alternative energy carriers, hydrogen has attracted significant attention due to its high energy density, renewability, and zero carbon emission characteristics, making it a strong candidate for replacing conventional fossil fuels \cite{lubitz2007hydrogen}. Hydrogen has an exceptionally high gravimetric energy density of 142 MJ/kg, which is nearly three times higher than that of gasoline ($\sim$47 MJ/kg), highlighting its potential as an efficient and sustainable energy source \cite{barbir1992hydrogen, song2024optimizing}. Despite these advantages, the practical application of hydrogen energy remains significantly constrained by the lack of an efficient, safe, and economically viable storage system.

Hydrogen can be produced through multiple pathways, including photocatalytic water splitting \cite{preethi2013photocatalytic}, electrocatalysis \cite{gong2022perspective}, biological production \cite{balat2009production}, and high-temperature pyrolysis \cite{li2014investigation}. However, efficient storage and transportation remain the primary bottlenecks preventing widespread adoption of hydrogen. Conventional hydrogen storage methods involve either high-pressure gas compression (up to 700 bar) or cryogenic liquefaction (at 20 K). These approaches suffer from high energy consumption, safety hazards, limited storage capacities, and significant economic costs \cite{schlapbach2001hydrogen}. Given these challenges, solid-state hydrogen storage has emerged as a promising alternative due to its advantages in safety, reversibility, and cost-effectiveness \cite{jena2011materials, liu2020trends}. This method involves the adsorption of hydrogen into solid materials, allowing for the controlled release of hydrogen under moderate conditions.

To facilitate the integration of hydrogen into commercial applications, the U.S. Department of Energy (DOE) has set ambitious targets for on-board hydrogen storage systems, requiring materials to achieve a gravimetric capacity of at least 6.5 wt\% by 2025 for light-duty fuel cell applications \cite{garcia2018benchmark}. Furthermore, an optimal hydrogen adsorption energy between -0.1 eV/H$_2$ and -0.4 eV/H$_2$ is essential to ensure both stability and reversibility under ambient conditions \cite{abe2019hydrogen}. Consequently, extensive research has been dedicated to identifying materials capable of meeting these stringent requirements. Various materials such as metal hydrides \cite{ZHONG2025148, LI2025109566, WANG2025159591, GUEMOU202483}, metal-organic frameworks (MOFs) \cite{rosi2003hydrogen, langmi2014hydrogen}, zeolites \cite{martin2018identifying, ramirez2007dihydrogen}, MXenes \cite{kumar2021mxenes, shahzad20202d, https://doi.org/10.1002/adfm.202418230}, nanocages \cite{WANG20143780}, and carbon-based materials \cite{sun2006first, cheng2001hydrogen} have been explored for hydrogen storage. Among these, two-dimensional (2D) materials have gained significant attention due to their tunable electronic properties, large surface-to-volume ratio, and high adsorption efficiency \cite{gopalsamy2014hydrogen, jaiswal2022alkali}. The functionalization of 2D materials with alkali, alkaline earth, and transition metals has been shown to significantly enhance their hydrogen adsorption capabilities by optimizing the binding energies within the desired range.

A particularly promising approach involves incorporating superalkali clusters, such as OLi$_2$, OLi$_3$, and NLi$_4$, into 2D materials \cite{zhang2024enhanced}. Superalkali clusters exhibit unique properties, including low ionization energy and strong electron-donating behavior, which enhance the adsorption of hydrogen molecules via electrostatic interactions. Their incorporation into 2D substrates modulates the charge distribution, effectively strengthening the hydrogen binding energy while maintaining reversibility under ambient conditions \cite{you2024dft}. Recent studies have demonstrated the feasibility of using superalkali-functionalized materials for hydrogen storage. For example, OLi-functionalized hexagonal boron nitride (h-BN) has been reported to adsorb multiple hydrogen molecules with notable adsorption energies, facilitating stable storage \cite{naqvi2017hexagonal}. Similarly, OLi$_2$-decorated h-BN has achieved impressive hydrogen storage densities, highlighting its potential as a lightweight and efficient hydrogen carrier \cite{hussain2013hexagonal}. Other investigations have shown that NLi$_4$ clusters anchored on haeckelite structures can significantly improve hydrogen storage capacity, with adsorption energies and gravimetric densities exceeding many conventional storage materials \cite{chen2023reversible, zhang2022hydrogen}. Therefore, superalkali cluster decoration exhibits remarkable potential for hydrogen storage, effectively addressing existing limitations and enhancing performance.

Among the various 2D carbon-based materials, irida-graphene (IG) has recently been identified as a promising candidate for hydrogen storage due to its unique 3-6-8 ring structure, which offers enhanced electronic properties and binding sites for functionalization \cite{junior2023irida}. However, pristine IG, like most carbon-based materials, exhibits weak van der Waals interactions with hydrogen molecules, limiting its storage capacity. To address this limitation, previous studies have investigated the decoration of IG with metals such as Ti, Li, Na, and K to enhance hydrogen adsorption. For instance, Ti-decorated IG was shown to adsorb up to five H$_2$ molecules per Ti atom, achieving a gravimetric hydrogen density of 7.7 wt\% with an average adsorption energy of -0.41 eV/H$_2$, ensuring stability at temperatures as high as 600 K \cite{TAN2024738}. Similarly, Li-functionalized IG exhibited a hydrogen storage capacity of 7.06 wt\% with adsorption energies ranging from -0.230 eV/H$_2$ to -0.276 eV/H$_2$ \cite{ZHANG20241004}. Furthermore, studies on IG modified with alkali metals revealed that Na-IG systems achieved a storage capacity of 7.82 wt\%, surpassing the DOE target, with strong binding stability and moderate desorption temperatures \cite{YUAN2024114756, DUAN20241}.

Based on these promising developments, this study explores the functionalization of IG with superalkali OLi$_3$ clusters for enhanced hydrogen storage. Using density functional theory (DFT) calculations, we systematically investigate the adsorption energy, hydrogen storage capacity, and desorption behavior of OLi$_3$-decorated IG. In addition, \textit{ab initio} molecular dynamics (AIMD) simulations are employed to assess the thermal stability and reversibility of hydrogen adsorption. The objective of this study is to evaluate whether OLi$_3$-functionalized IG can surpass conventional alkali metal modifications in terms of hydrogen adsorption efficiency and desorption kinetics. Our findings provide key insights into the feasibility of superalkali-functionalized 2D materials for next-generation hydrogen storage applications, potentially contributing to the realization of sustainable hydrogen-based energy systems.

\section{Computational Setup}

This work used the generalized gradient approximation (GGA) \cite{PhysRevLett.77.3865} based on the Perdew, Burke, and Ernzerhof (PBE) \cite{ernzerhof1999assessment} functional and augmented wave (PAW) method \cite{PhysRevB.50.17953}, both implemented in the Vienna ab initio simulation package (VASP). A cutoff energy of 520 eV was utilized to represent the electronic states. To prevent interactions between periodic adjacent layers, a vacuum layer of 15 Å was fixed along the c-axis. A $\Gamma$-centered k-point grid of 5 $\times$ 5 $\times$ 1 k-point grid was used for structural optimization to balance computational cost and accuracy. A denser 9 $\times$ 9 $\times$ 1 grid was employed for accurate electronic structure calculations, including density of states and band structures. The DFT-D2 dispersion correction proposed by Grimme \cite{grimme2010consistent} was included. Structural optimization was performed using the conjugate gradient algorithm until the convergence criteria were met, requiring the energy errors for the atomic positions and the lattice parameters to be less than $1 \times 10^{-5}$ eV and the Hellmann-Feynman forces on each atom to be within 0.01 eV/\AA. To evaluate the thermodynamic stability of the OLi$_3$-decorated IG and its hydrogen storage reversibility, \textit{ab initio} molecular dynamics (AIMD) \cite{martyna1992nose} simulations were performed. Furthermore, the charge transfer was quantified by Bader charge analysis \cite{bader1985atoms}.

The hydrogen adsorption energy (E$_{\text{ads}}$) for the OLi$_3$@IG + nH$_2$ configurations is determined using the equation \cite{BENIWAL2025114947}: 

\begin{equation}
E_{\text{ads}} = \frac{1}{n} \left( E_{\text{total}} - E_{\text{sub}} - nE_{\text{H}_2} \right)
\end{equation}

where $E_{\text{total}}$ is the total energy of the OLi$_3$@IG + nH$_2$ system, $E_{\text{sub}}$ refers to the energy of the OLi$_3$@IG substrate, and $E_{\text{H}_2}$ is the energy of an isolated H$_2$ molecule.

The hydrogen adsorption capacity (HAC) is defined as \cite{elaggoune2024stability}: 

\begin{equation}
HAC = \frac{n_H M_H}{n_C M_C + n_O M_O + n_{\text{Li}} M_{\text{Li}} + n_H M_H}
\end{equation}

where $n_X$ and $M_X$ are the number of atoms and molar masses of H, C, O, and Li, respectively.

The hydrogen release temperature (T$_{\text{R}}$) is obtained using the van't Hoff equation \cite{alhameedi2019metal}, considering atmospheric pressure (1 atm):

\begin{equation}
T_{\text{R}} = \left| E_{\text{ads}} \right|\frac{R}{K_B \Delta S}
\end{equation}

where $R$ and $k_B$ are the universal gas constant and the Boltzmann constant, and $\Delta S$ is the change in entropy of hydrogen from gas to liquid phase (75.44 J mol$^{-1}$ K$^{-1}$).

An analysis based on thermodynamics was conducted to assess how H$_2$ molecules adsorb and desorb under real-world conditions, employing the grand canonical partition function \(Z\), described by:
\begin{equation}
   Z = 1 + \sum_{i=1}^{n} \exp\left(\frac{E_i^{\text{ads}} - \mu}{k_B T}\right)
\end{equation}
in this equation, \(n\) signifies the total number of H$_2$ molecules potentially adsorbed, while $\mu$ signifies the chemical potential for the H$_2$ molecule in the gaseous state. Here, $E_{i}^{ads}$ represents the energy required to adsorb the $i^{th}$ H$_2$ molecule. The methodology and detailed calculations of this study have previously been documented in previous studies \cite{hashmi2017ultra, kaewmaraya2023ultrahigh}.

\section{Results and Discussion}

\begin{figure*}[!ht]
    \centering
    \includegraphics[width=0.8\linewidth]{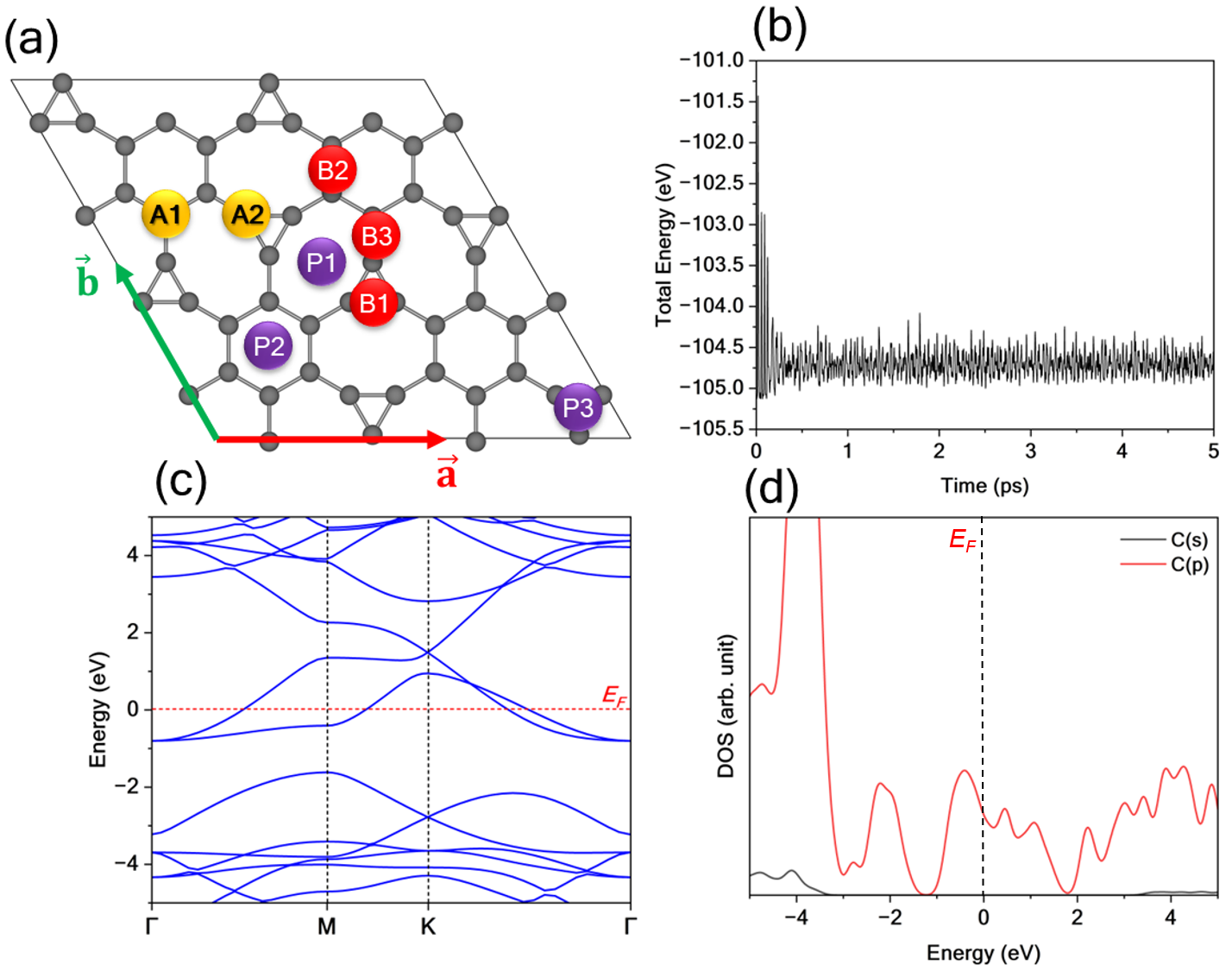}
    \caption{(a) Representative unit cell of IG and its high-symmetry sites considered for OLi$_3$-decoration; (b) energy fluctuations from AIMD simulations at 300 K; (c) electronic band structure; and (d) projected density of states (PDOS) of pristine IG. The dashed lines at 0 eV for (a) and (b) denote the Fermi level (E$_F$).}
    \label{fig:unit-cell}
\end{figure*}

IG monolayer is characterized by a hexagonal structure (No. 191) with lattice parameters $a = b =$ 6.32~\AA, as depicted in Figure~\ref{fig:unit-cell}a. This structure consists of two non-equivalent carbon atoms, $C_1$ (0.741, 0.481, 0.000) and $C_2$ (0.869,  0.737,  0.000), where the values in parentheses indicate their internal coordinates. These results are consistent with previous reports \cite{ZHANG20241004, DUAN20241, YUAN2024114756}.

To assess the thermal stability of the IG monolayer, AIMD simulations were performed for 5 ps at 300 K, as shown in Figure~\ref{fig:unit-cell}b. The total energy remains stable throughout the simulation, exhibiting minor fluctuations of less than 0.5 eV after equilibrium.

The electronic properties of IG were analyzed through its band structure and projected density of states (PDOS), presented in Figures~\ref{fig:unit-cell}c and \ref{fig:unit-cell}d. The results demonstrate the metallic nature of IG, with two bands crossing the Fermi level ($E_F$) and a cone-shaped crossing appearing at the K-point. The PDOS analysis reveals a predominant contribution from $p$-states near $E_F$, attributed to the strong $sp^2$ hybridization within the monolayer. This hybridization improves the band dispersion closer to $E_F$, further reinforcing the conductor behavior of IG.

In order to thoroughly analyze OLi$_3$ decoration on the IG monolayer, eight distinct high-symmetry sites were taken into account, as illustrated in Figure 1a. These sites can be categorized as atom sites (A1 and A2), pore sites (P1, P2, and P3), and bond sites (B1, B2, and B3). Table \ref{tbl2} provides the adsorption energies ($E_{ads}$) corresponding to each site, together with the optimized final positions. During analysis, OLi$_3$ was initially placed 2~\AA~above the IG monolayer at each site, subsequently allowing the system to optimize without any constraints on its geometry. It is observed that $E_{ads}$ ranges between -3.13 eV and -3.24 eV, which are considerably lower (or higher in absolute terms) than the experimentally determined cohesive energy of the bulk Li, -1.63 eV \cite{PhysRevB.6.3637, JIANG2024865}. In particular, the B3 and P3 sites exhibit nearly identical $E_{ads}$ values, demonstrating close degeneracy, with the lowest adsorption energies recorded as -3.24 eV and -3.23 eV, respectively. Our results align with those reported by You et al. \cite{you2024dft}, and Beniwal and Kumar \cite{beniwal2025superalkali} that observed a $E_{ads}$ of -3.82 eV and -3.05 eV for Graphyne, and biphenylene sheet, respectively. Therefore, the negative obtained $E_{ads}$ confirms the thermodynamic favorability of the interaction. In view of the above-mentioned, the B3 site was selected to conduct the OLi$_3$ decoration on the IG monolayer. 

\begin{table}[!ht]
\centering
\caption{Adsorption energies (E$_\text{ads}$) and optimized configurations of the adsorption sites evaluated for OLi$_3$ decoration on irida-graphene (IG).}
\label{tbl2}
\begin{tabular*}{\linewidth}{@{\extracolsep{\fill}} lcc}
\toprule
\textbf{Initial site} & \textbf{E$_\text{ads}$ (eV)} & \textbf{Final site} \\
\midrule
A1 &  -3.24  & B3  \\
A2 &  -3.23  & P3  \\
B1 &  -3.23  & P3  \\
B2 &  -3.20  & Between B2 and P1   \\
B3 &  -3.24  & A2  \\
P1 &  -3.19  & Between B2 and P1  \\
P2 &  -3.13  & P2  \\
P3 &  -3.23  & P3  \\
\bottomrule
\end{tabular*}
\end{table}

\begin{figure*}[!ht]
    \centering
    \includegraphics[width=0.8\linewidth]{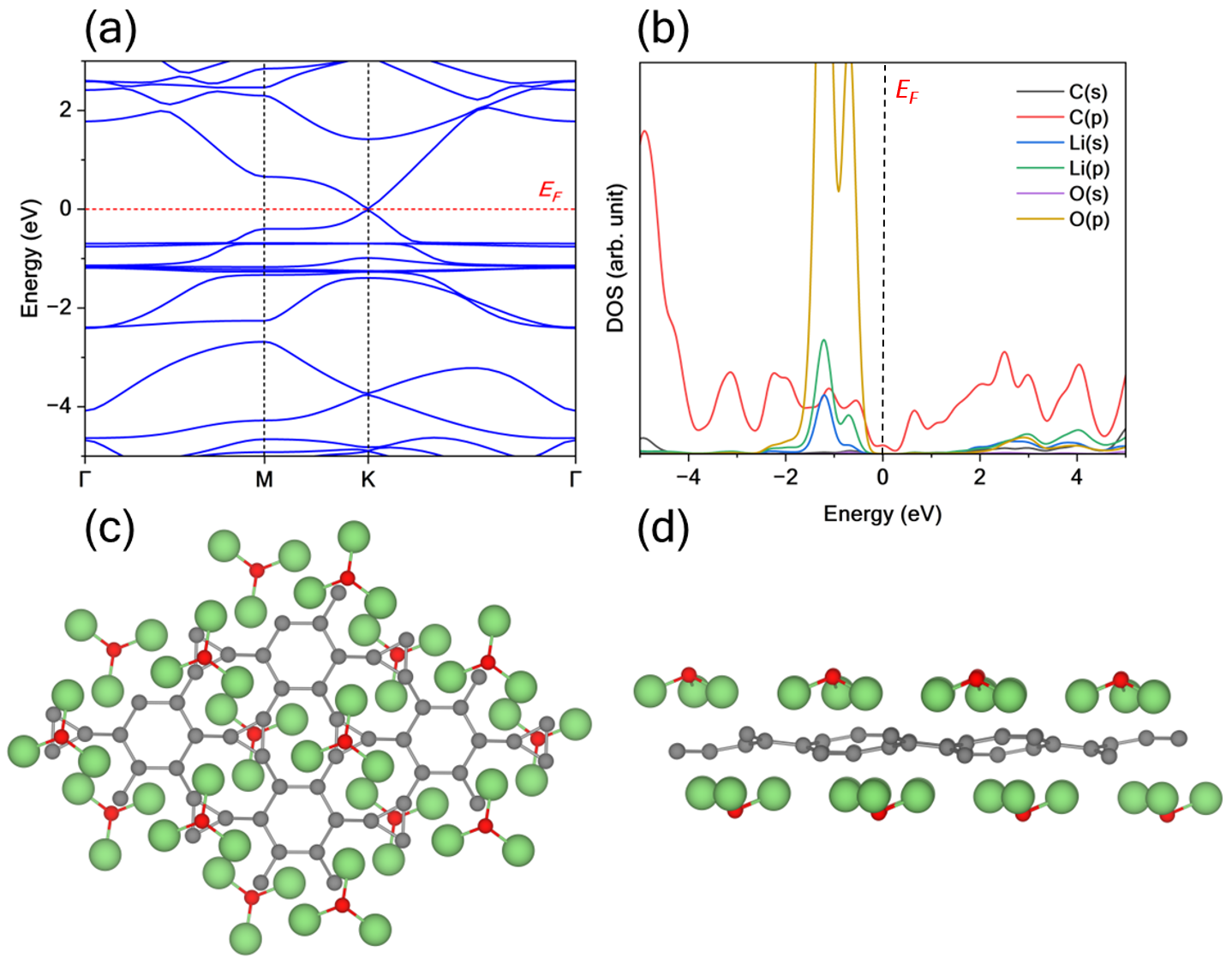}
    \caption{(a) Band structure and (b) projected density of states (PDOS) of the OLi$_3$@IG system. (c) Top and (d) side views of the optimized structure. Grey, green, and red spheres represent C, Li, and O atoms, respectively. The dashed lines at 0 eV for (a) and (b) denote the Fermi level (E$_F$).}
    \label{fig:OLi3_decoration}
\end{figure*}

Due to the minimal energy disparity between the B3 and P3 sites, the optimized configuration resulted in OLi$_3$ clusters placed between these two sites, as illustrated in Figure \ref{fig:OLi3_decoration}. This figure also displays the band structure, PDOS, and top and side views of the optimized OLi$_3$@IG system. Electronically, it is notable that the $E_F$ is elevated compared to the pristine IG, indicating that previously unoccupied states in the IG become occupied due to the interaction with OLi$_3$ clusters, which have moved toward the Dirac cone. PDOS reveals a notable contribution from the OLi$_3$ states near the $E_F$, suggesting a significant overlap with the delocalized p states of both IG and OLi$_3$. The Bader charge analysis further supports the occupation of previously empty IG states, quantifying a charge transfer of -0.947 |e|/OLi$_3$ from the OLi$_3$ clusters to IG.

Furthermore, Figures \ref{fig:OLi3_decoration}c and \ref{fig:OLi3_decoration}d demonstrate that incorporation of OLi$_3$ clusters induces a buckling effect in the IG structure, with no reconstructions or substantial distortions observed. A further prominent aspect when a substrate is functionalized is to note the metal-to-metal distance, or, in this case, the distances between OLi$_3$ clusters, which are essential to avoid any molecular aggregation. Here, the minimum obtained OLi$_3$-OLi$_3$ distance is found to be 3.92~\AA, which indicates that the OLi$_3$ molecules are not sufficiently close to have interactions between them. Furthermore, the optimized OLi$_3$ cluster exhibits an O-Li bond distance of 1.67~\AA, which increases slightly to 1.70~\AA~in the OLi$_3$@IG system, indicating a stable adsorption configuration.

\begin{figure}[!ht]
    \centering
    \includegraphics[width=0.8\linewidth]{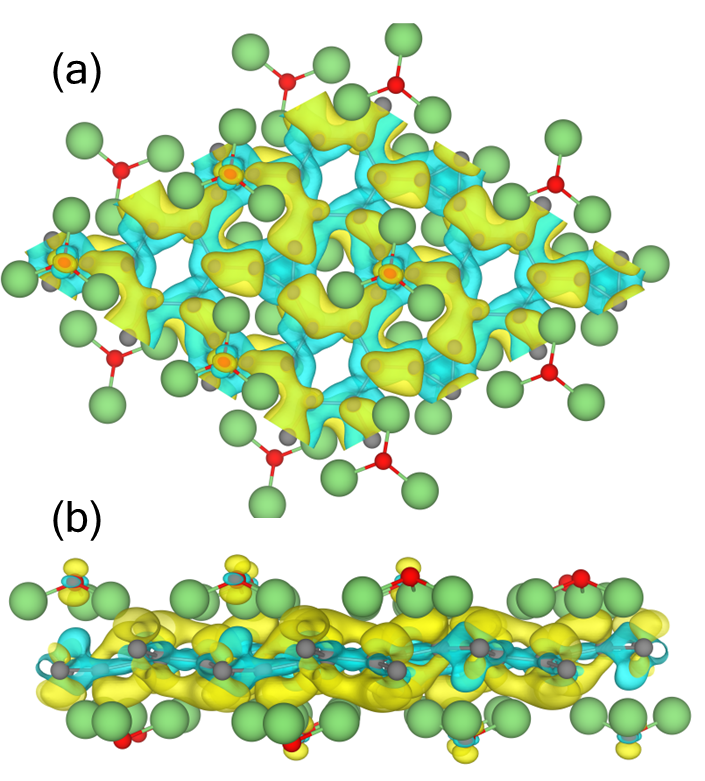}
    \caption{(a) Top and (b) side views of the charge density difference (CDD) map for the OLi$_3$@IG system. Yellow and blue regions indicate charge accumulation and depletion, respectively. Grey, green, and red spheres represent C, Li, and O atoms, respectively.}
    \label{fig:CDD_OLi3}
\end{figure}

To evaluate the charge redistribution in the OLi$_3$@IG complex, the charge density difference (CDD) map was calculated and presented in Figure~\ref{fig:CDD_OLi3}. These maps illustrate the regions of charge accumulation (depicted in yellow) and charge depletion (depicted in blue), providing insights into the electronic interactions between IG and OLi$_3$. A noticeable charge depletion is observed along the bond axes of the IG monolayer, while charge accumulation occurs predominantly above these axes. This charge redistribution suggests that the $\pi$ states of IG play a crucial role in facilitating charge transfer from the OLi$_3$ cluster to the IG monolayer. This behavior is indicative of a strong electronic coupling between IG and OLi$_3$.

\begin{figure}[!ht]
    \centering
    \includegraphics[width=1\linewidth]{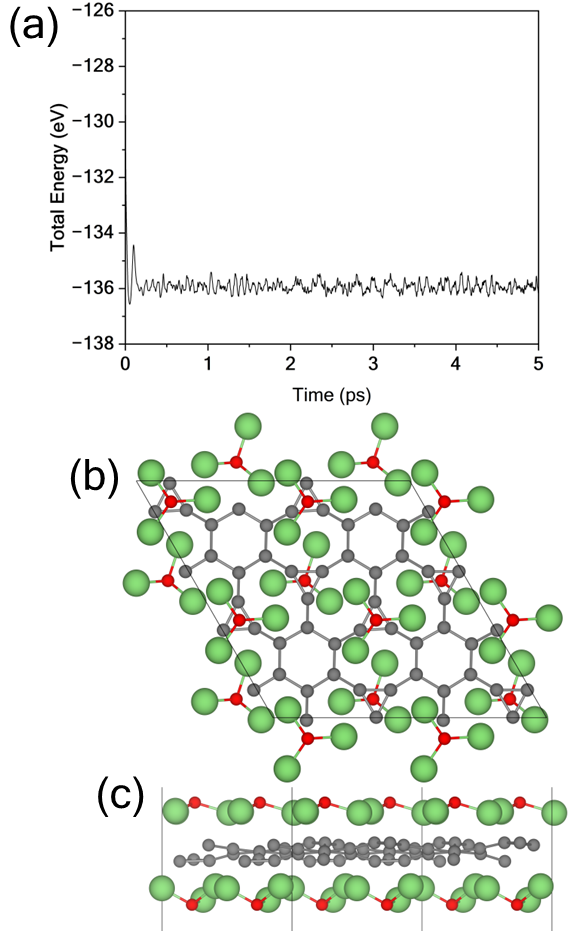}
    \caption{(a) Energy fluctuations during AIMD simulations at 300 K over 5 ps, along with the (b) top and (c) side views of the OLi$_3$@IG structure. Grey, green, and red spheres represent C, Li, and O atoms, respectively.}
    \label{fig:AIMD_OLi3}
\end{figure}

Figure \ref{fig:AIMD_OLi3} illustrates the AIMD simulations conducted at 300 K over 5 ps to assess the thermal stability of the OLi$_3$@IG complex. The energy displayed minor fluctuations, confirming the absence of phase transitions or desorption of the OLi$_3$ clusters. Furthermore, the final structure reveals the intactness of the OLi$_3$@IG framework, without bond rearrangements, which denotes the complex as stable at room temperature. The OLi$_3$ clusters persist at their original positions at the end of the simulations, underscoring their stable adsorption.

\begin{figure*}[!ht]
    \centering
    \includegraphics[width=0.8\linewidth]{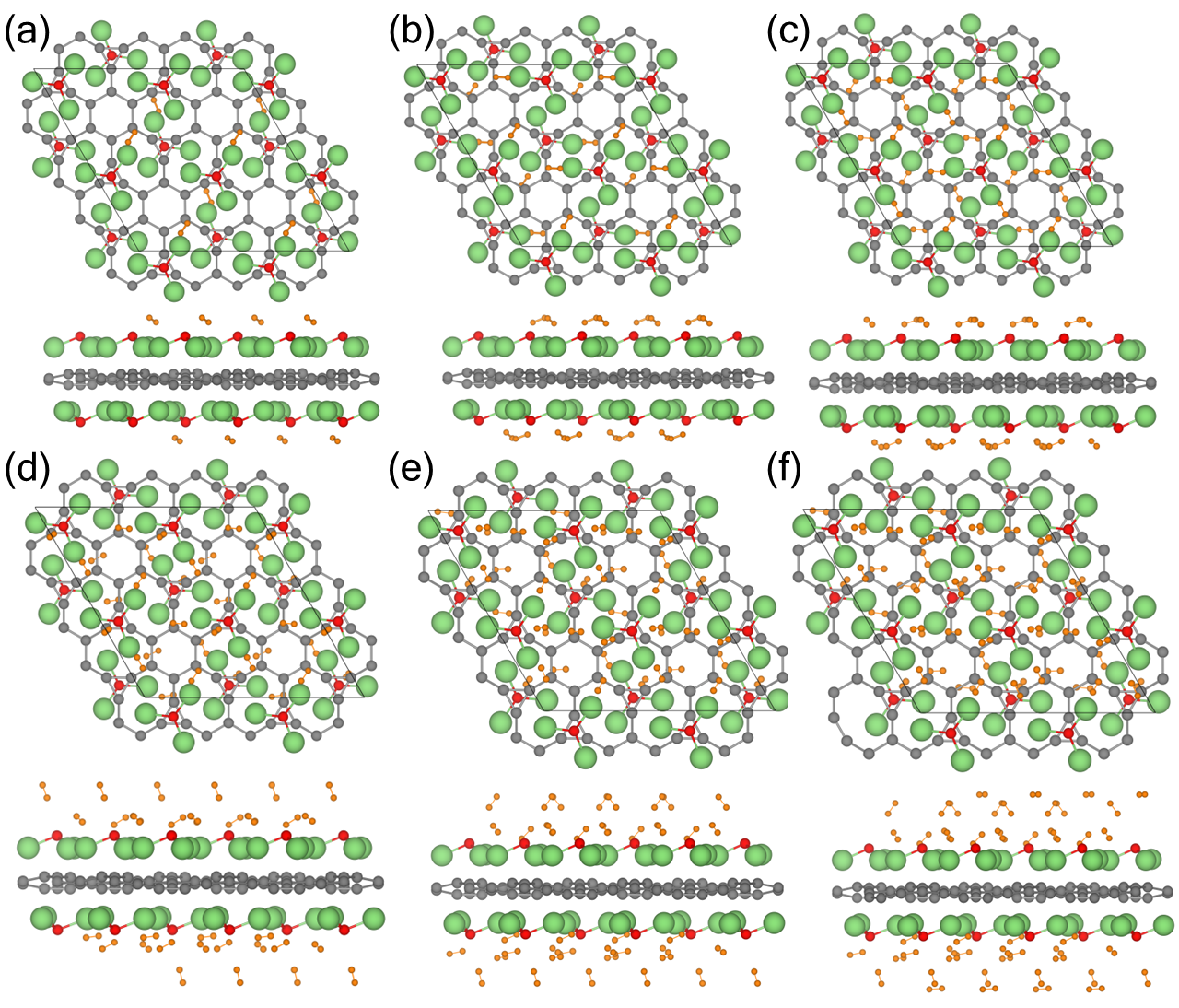}
    \caption{Optimized structures of OLi$_3$@IG with adsorbed H$_2$ molecules: (a) 2H$_2$, (b) 4H$_2$, (c) 6H$_2$, (d) 8H$_2$, (e) 10H$_2$, and (f) 12H$_2$. The stoichiometry corresponds to the IG unit cell. Grey, green, red, and orange spheres represent C, Li, O, and H atoms, respectively.}
    \label{fig:saturation}
\end{figure*}

To investigate hydrogen storage in the IG@OLi$3$ system, H$_2$ molecules were sequentially adsorbed, as illustrated in Figure~\ref{fig:saturation}. The results for each configuration, including the $E_{ads}$, HAC, the average bond length of OLi$_3$ clusters (R$_\text{O-Li}$) and H$_2$ molecules (R$_\text{H-H}$), and the T$_\text{R}$, are summarized in Table~\ref{tbl1}. From a structural perspective, it has been confirmed that the inclusion of H$_2$ molecules does not lead to any notable displacement of OLi$_3$ or distortion of bonds within the IG monolayer. This observation aligns with the inherent characteristics of the H$_2$ molecules and the OLi$_3$@IG substrate, which are associated with weak physisorption, as can be seen in the following.

Firstly, the $E_{ads}$ varies between -0.27 eV and -0.19 eV, indicating a weak physisorption interaction between the H$_2$ molecules and the OLi$_3$@IG substrate. These values fall within the desirable range for reversible hydrogen storage (between -0.1 eV and -0.4 eV). Regarding HAC, the system achieves a maximum adsorption configuration of 12 H$_2$ molecules per OLi$_3$@IG unit cell, corresponding to an impressive storage capacity of 10.00~wt\%. In Table~\ref{comparison} it is possible to compare the properties of OLi$_3$@IG with other IG-based and other relevant systems recently reported. OLi$_3$@IG has an adsorption energy of 0.19 eV (in absolute value) per H$_2$, which position it within the optimal range for effective hydrogen storage. This value is comparable to Ca@IG (0.18 eV) and slightly lower than Li@IG (0.28 eV) and Ti@IG (0.41 eV), which exhibit stronger adsorption, but may lead to difficulties in desorption due to excessive binding energy. However, Na@IG (0.14 eV) and Na@IGP-SiC (0.10 eV) show weaker adsorption, which may not be sufficient to retain hydrogen under practical conditions.

Regarding HAC, OLi$_3$@IG demonstrates the highest capacity among the systems analyzed, reaching 10.00 wt\%. This surpasses Li@IG (7.06 wt\%), Na@IG (7.82 wt\%), Ca@IG (8.00 wt\%) and Ti@IG (7.70 wt\%). Besides that, one can note that our hydrogen capacity is superior than other 2D materials under OLi$_3$ decoration, such as Biphenylene (9.11 wt\%) and h-BN (9.67 wt\%). The significantly higher HAC of OLi$_3$@IG highlights its superior ability to store hydrogen efficiently, making it an attractive candidate for practical applications.

The $T_{\text{R}}$ of OLi$_3$@IG + 12H$_2$ is 237.74 K, which falls within a reasonable range for hydrogen release under controlled conditions. It is higher than Na@IGP-SiC (148 K) and Li@IGP-SiC (191 K), indicating improved hydrogen retention. However, it remains lower than Li@IG (353 K) and Ti@IG (524 K), which may require elevated temperatures to complete hydrogen release, potentially affecting energy efficiency.

In terms of structural stability, the interaction with the OLi$_3$@IG substrate does not significantly affect the integrity of the H$_2$ molecules, as evidenced by a minimal bond length variation of just 0.01~\AA. Finally, the values of $T_\text{R}$, ranging from 339.61 K (for 2 H$_2$ molecules) to 237.74 K (for 12 H$_2$ molecules), indicate that hydrogen can be easily released under ambient conditions, further confirming the feasibility of OLi$_3$@IG as a promising material for reversible hydrogen storage.

\begin{table*}[!ht]
\caption{Adsorption energy (E$_\text{ads}$), hydrogen adsorption capacity (HAC), average O--Li (R$_\text{O-Li}$) and H--H (R$_\text{H-H}$) bond lengths, and release temperature (T$_\text{R}$) for OLi$_3$@IG + nH$_2$  (n = 2, 4, 6, 8, 10, and 12).}
\label{tbl1}
\centering
\begin{tabular*}{\linewidth}{@{\extracolsep{\fill}} lccccc @{}}
\toprule
\textbf{System} & \textbf{E$_\text{ads}$ (eV)} & \textbf{HAC (wt\%)} & \textbf{R$_\text{O-Li}$ (\AA)} &  \textbf{R$_\text{H-H}$ (\AA)} & \textbf{T$_\text{R}$ (K)} \\
\midrule
\textbf{OLi$_3$@Irida-graphene + 2H$_2$}    & -0.27   & 1.82 & 1.71 & 0.76 & 339.61 \\
\textbf{OLi$_3$@Irida-graphene + 4H$_2$}    & -0.27   & 3.57 & 1.72 & 0.77 & 341.07 \\
\textbf{OLi$_3$@Irida-graphene + 6H$_2$}    & -0.27   & 5.26 & 1.72 & 0.77 & 340.90 \\
\textbf{OLi$_3$@Irida-graphene + 8H$_2$}    & -0.23  & 6.90 & 1.72 & 0.76 & 288.49 \\
\textbf{OLi$_3$@Irida-graphene + 10H$_2$}    & -0.21  & 8.47 & 1.73 & 0.76 & 268.03 \\
\textbf{OLi$_3$@Irida-graphene + 12H$_2$}    & -0.19  & 10.00 & 1.73 & 0.76 & 237.74 \\
\bottomrule
\end{tabular*}
\end{table*}

\begin{table}[!ht]
\centering
\caption{Absolute adsorption energy per H$_2$ (|E$_\text{ads}$|), Hydrogen Adsorption Capacity (HAC), and Desorption temperature (T$_\text{des}$) (at 1 atm) associated with configurations exhibiting complete H$_2$ coverage configurations in IG-based and other relevant systems.}
\label{comparison}
\begin{tabular*}{\linewidth}{@{\extracolsep{\fill}} lccc @{}}
\toprule
\textbf{System}                      & \textbf{|E$_\text{ads}$| (eV)} & \textbf{HAC (wt\%)} & \textbf{T$_\text{des}$ (K)} \\
\midrule
\textbf{OLi$_3$@IG} & 0.19 & 10.00 & 238 \\
\textbf{Li@IG} \cite{ZHANG20241004} & 0.28 & 7.06 & 353 \\
\textbf{Na@IG} \cite{DUAN20241} & 0.14 & 7.82 & 248 \\
\textbf{Ti@IG} \cite{TAN2024738} & 0.41  & 7.70 & 524  \\
\textbf{Ca@IG} \cite{ZHANG20241004} & 0.18  & 8.00 & 226 \\
\textbf{OLi$_3$@BPN}\cite{beniwal2025superalkali} & 0.28 & 9.11 & 258 \\
\textbf{OLi$_3$@h-BN}\cite{zhang2023potential} & 0.18 & 9.67 & 234 \\
\textbf{Na@B$_7$N$_5$} \cite{LIU2025105802} & 0.20  & 7.70  & 257  \\ 
\textbf{Li@B$_5$N$_3$} \cite{doi:10.1021/acs.langmuir.4c00779}  & 0.21  & 6.30  & 267 \\ 
\textbf{Li@net-Y} \cite{CHEN2024114445} & 0.27 & 9.00 & 348 \\
\textbf{Li@$\alpha$-C$_3$N$_2$} \cite{CHEN2024510} & 0.22 & 5.70 & 285 \\
\textbf{Li@IGP-SiC} \cite{MARTINS202498} & 0.14 & 8.27 & 191 \\
\textbf{Na@IGP-SiC} \cite{MARTINS202498} & 0.10 & 6.78 & 148 \\
\bottomrule
\end{tabular*}
\end{table}

\begin{figure}
    \centering
    \includegraphics[width=0.8\linewidth]{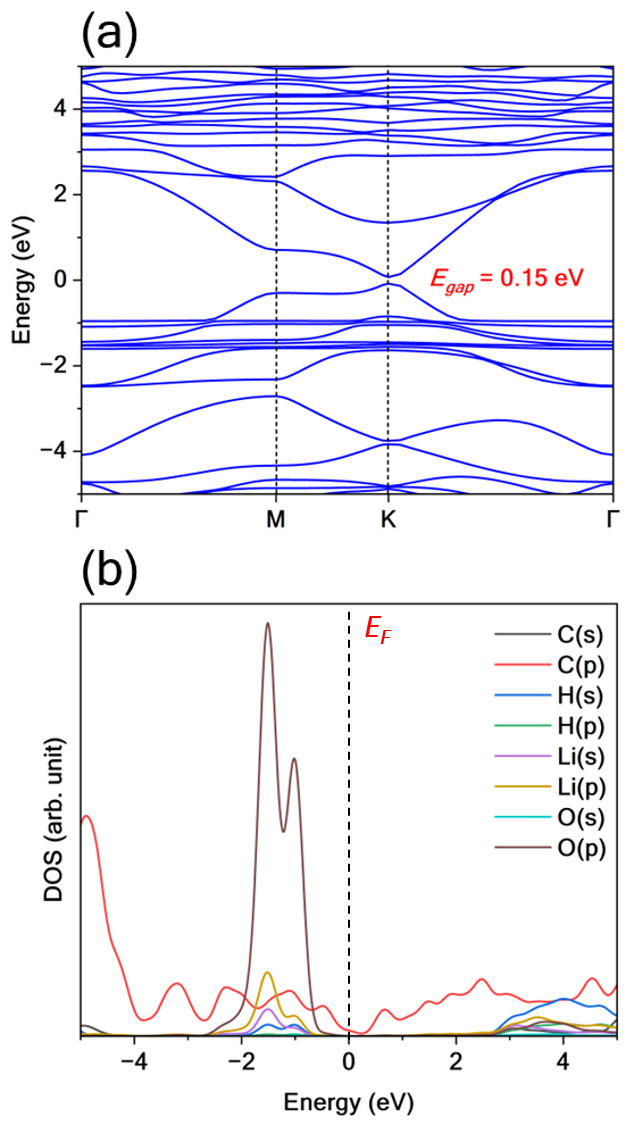}
    \caption{(a) Band structure and (b) projected density of states (PDOS) of the OLi$_3$@IG + 12H$_2$ system. The system exhibits semiconducting behavior, with the Fermi level (E$_F$) set at 0 eV.}
    \label{fig:12H2}
\end{figure}

Figure \ref{fig:12H2} illustrates the band structure and PDOS of OLi$_3$@IG + 12H$_2$, offering insight into its electronic properties. The introduction of H$_2$ molecules induces an opening in the band gap, characterized by a direct K-point transition. The calculated band gap energy, E$_{gap}$, is 0.15 eV. Notably, the valence band maximum (VBM) and the conduction band minimum (CBM) nearly mirror each other, and the bands exhibit significant dispersion. PDOS reveals that at energies near the CBM, the H$_2$ and OLi$_3$ states demonstrate strong coupling, with PDOS curves showing similar shapes. Overall, PDOS is predominantly composed of the C(p) states, although the O(p) and Li(p) states contribute more significantly from -2.5 to -0.5 eV. For conduction band energies exceeding 3 eV, substantial contributions from the OLi$_3$ and H$_2$ states are observed. Our results suggest that OLi$_3$@IG can serve as a platform for hydrogen storage and detection applications, as changes in its electronic response upon gas exposure indicate strong sensing applications.

\begin{figure}[!ht]
    \centering
    \includegraphics[width=0.8\linewidth]{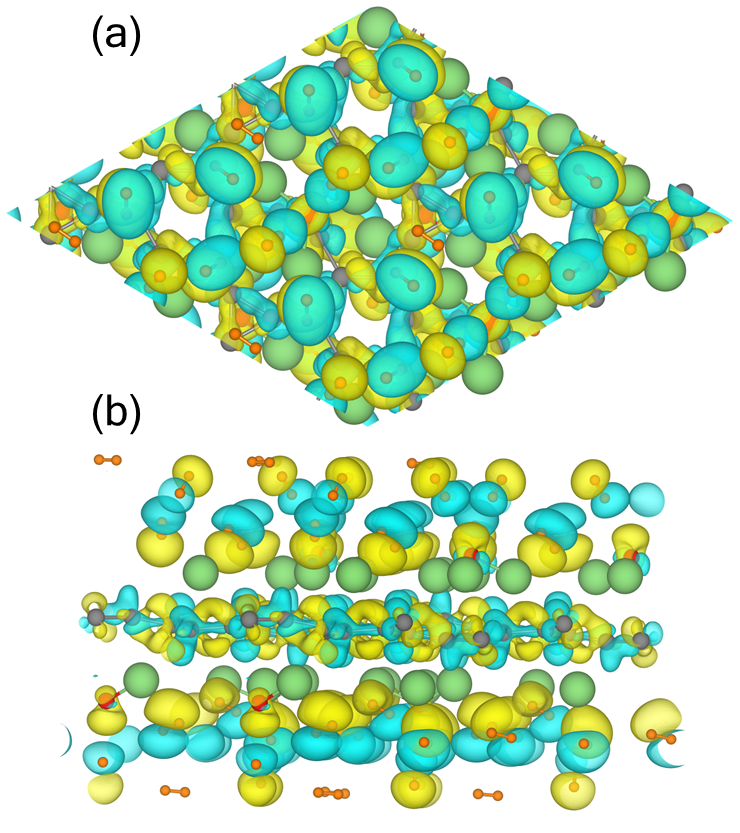}
    \caption{(a) Top and (b) side views of charge density difference (CDD) map for OLi$_3$@IG + 12H$_2$ system, where yellow (blue) denotes charge accumulation (depletion). Grey, green, red, and orange spheres indicate the C, Li, O, and H elements.}
    \label{fig:CDD_H2}
\end{figure}

For a complete understanding of the charge distribution within the OLi$_3$@IG + 12H$_2$ system, Figure \ref{fig:CDD_H2} illustrates the corresponding CDD maps. Observations reveal that there is no significant charge transfer between OLi$_3$@IG and H$_2$ when compared to the OLi$_3$@IG complex, as the CDD remains almost unchanged for IG and OLi$_3$. This is supported by Bader charge analysis, which assigns charge values of +0.01 |e|/H$_2$ and -0.01 |e|/H$_2$ to the OLi$_3$@IG and H$_2$ molecules, respectively. In particular, charge depletion and accumulation areas are present in the H$_2$ molecules, suggesting that charge polarization serves as the primary interaction mechanism between the OLi$_3$@IG and H$_2$ molecules.

\begin{figure*}[!ht]
    \centering
    \includegraphics[width=0.8\linewidth]{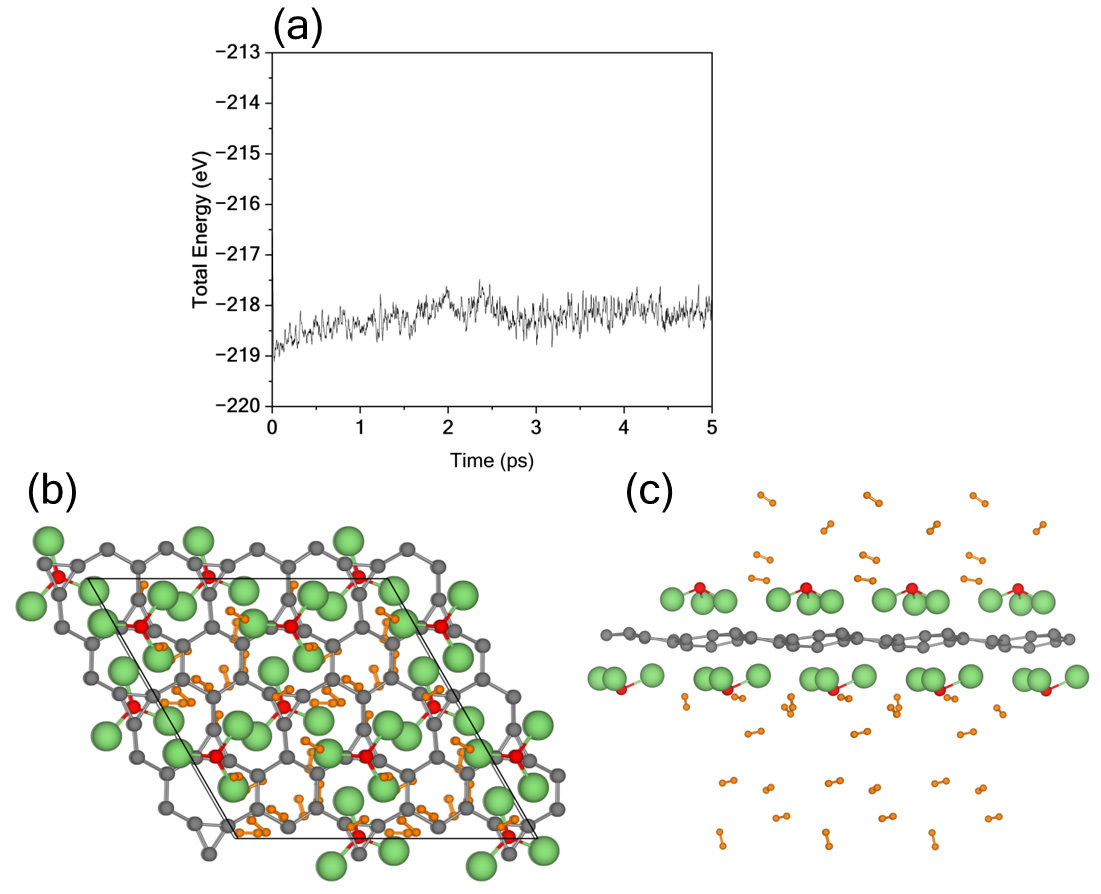}
    \caption{(a) Energetic profile for AIMD simulation, along with the (b) top and (c) side views of final structure for OLi$_3$@IG + 12H$_2$ system at 300 K. Grey, green, red, and orange spheres indicate the C, Li, O, and H elements.}
    \label{fig:MD+H2}
\end{figure*}

To analyze the thermal stability of the OLi$_3$@IG system under H$_2$ adsorption and desorption kinetics, the AIMD simulations were performed at 300 K considering an interval of 5 ps as represented by the energy fluctuations and the top and side views of the final structure in Figure \ref{fig:MD+H2}. It can be seen that the system exhibits lower energetic variations only after 2.5 ps, and irregular variations before 2.5 ps are related to H$_2$ molecules desorption events. Upon examination of the completed structure, it is apparent that a significant majority of H$_2$ molecules are liberated from the OLi$_3$@IG substrate. Throughout the process, the OLi$_3$@IG substrate maintains its structural integrity. The kinetics of beneficial desorption suggest advantages for reversible hydrogen storage, which favors the release of H$_2$ molecules at ambient room temperature.

\begin{figure}[!ht]
    \centering
    \includegraphics[width=1\linewidth]{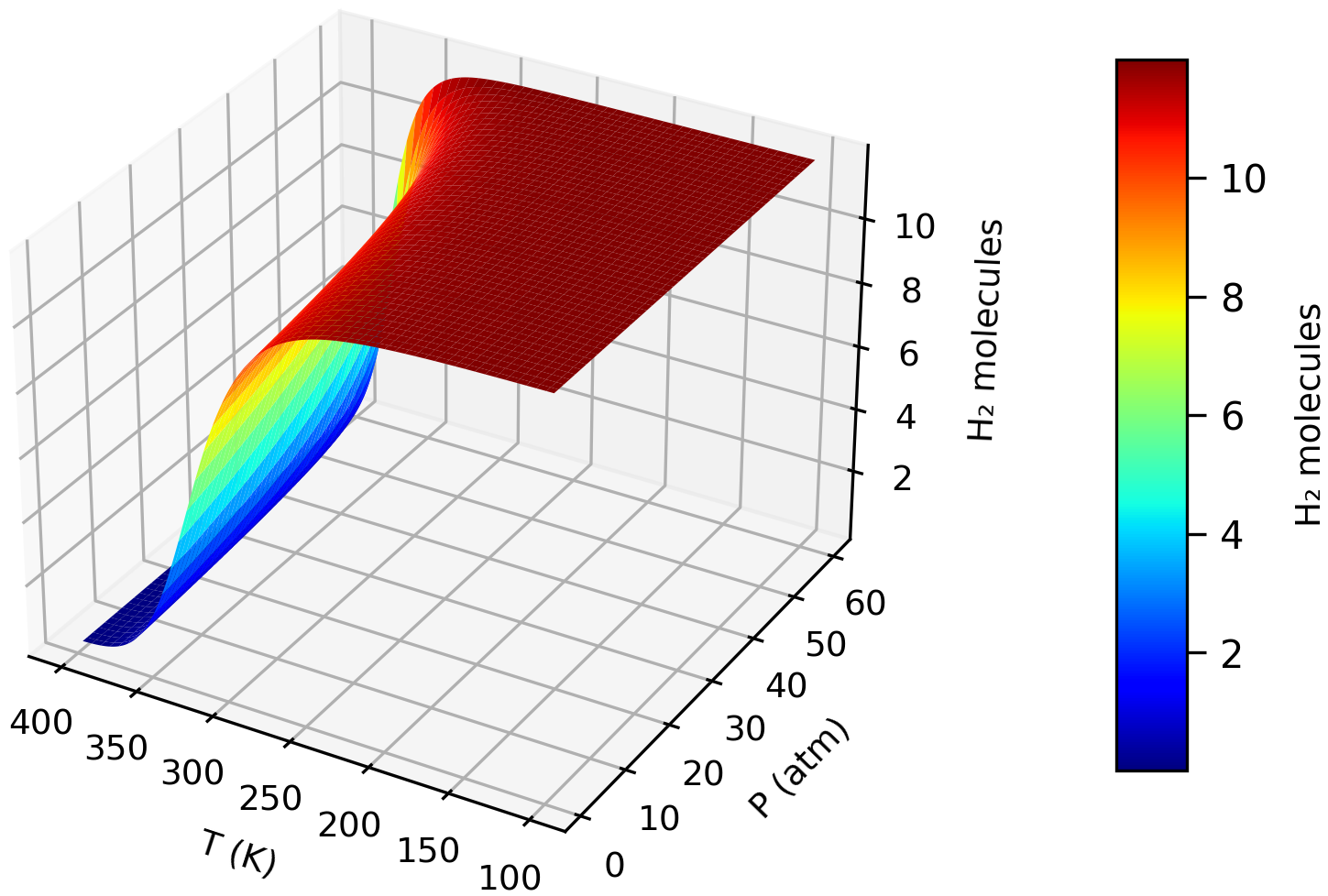}
    \caption{The average number of adsorbed H$_2$ on OLi$_3$@IG at various temperatures (T)
and pressures (P).}
    \label{fig:P_T}
\end{figure}

To evaluate the hydrogen storage performance of the OLi$_3$@IG substrate, the number of H$_2$ was analyzed for several values of pressure (P) and temperature (T) as represented in Figure \ref{fig:P_T}. In particular, it is interesting to analyze the adsorption (30 atm and 25$^\circ$C) and desorption (3 atm and 100$^\circ$C) conditions. The measurement of adsorption under the adsorption condition revealed that 11.56 molecules of H$_2$ were adsorbed, indicating an HAC of 9.64 wt\%. Upon reducing the pressure to 3 atm and increasing the temperature to 100$^\circ$C, nearly all adsorbed hydrogen molecules were released, with only 0.03 molecules remaining in the system. This result suggests that the interaction between H$_2$ and the material is predominantly physisorption, as no significant retention of hydrogen is observed after desorption.

\section{Conclusions}
This study investigates the potential of OLi$_3$-decorated irida-graphene (OLi$_3$@IG) for hydrogen storage through density functional theory (DFT) simulations.  Our results show that the OLi$_3$ cluster exhibits a significant binding energy of -3.24 eV when attached to one side of the IG on the most stable site, indicating a strong affinity between the OLi$_3$ cluster and the IG surface. The OLi$_3$@IG structure maintains its structural integrity at ambient temperature, exhibiting negligible distortions upon the introduction of hydrogen (H$_2$). The computed adsorption energy (E$_\text{ads}$) ranges from -0.27 eV to -0.19 eV, suggesting a minor decrease in adsorption strength with increasing number of hydrogen molecules. In addition, the release temperatures (T$_\text{R}$) decrease from 339.61 K to 237.74 K with higher hydrogen loading, indicating that increased hydrogen uptake facilitates desorption. The hydrogen storage capacity (HAC) shows a gradual increase, achieving up to 10.00 wt\% in the OLi$_3$@IG + 12H$_2$ configuration, underscoring its strong potential for hydrogen storage. Charge density difference (CDD) maps and Bader charge analysis revealed that charge polarization is the primary interaction mechanism between the OLi$_3$@IG and H$_2$ molecules. Ab initio molecular dynamics (AIMD) simulations reveal that H$_2$ molecules rapidly desorb from the OLi$_3$@IG + 12H$_2$ structure at 300 K, indicating favorable desorption kinetics.

\section*{Data access statement}
Data supporting the results can be accessed by contacting the corresponding author.

\section*{Conflicts of interest}
The authors declare that they have no conflict of interest.

\section*{Acknowledgements}
X.C. was funded by the Research Program of Chongqing Municipal Education Commission (No. KJQN202201327 and No. KJQN202301339) and Natural Science Foundation of Chongqing, China (CSTB2022NSCQ-MSX0621). This work was supported by the Brazilian funding agencies Fundação de Amparo à Pesquisa do Estado de São Paulo - FAPESP (grant no. 2022/03959-6, 2022/00349- 2, 2022/14576-0, 2020/01144-0, and 2022/16509-9), and National Council for Scientific, Technological Development - CNPq (grant no. 307213/2021–8).

\printcredits

\bibliography{cas-refs}

\begin{thebibliography}{10}

\bibitem{lenferna2018can}
Georges~Alexandre Lenferna.
\newblock Can we equitably manage the end of the fossil fuel era?
\newblock {\em Energy Research \& Social Science}, 35:217--223, 2018.

\bibitem{lubitz2007hydrogen}
Wolfgang Lubitz and William Tumas.
\newblock Hydrogen: an overview.
\newblock {\em Chemical reviews}, 107(10):3900--3903, 2007.

\bibitem{barbir1992hydrogen}
F~Barbir et~al.
\newblock Hydrogen: the wonder fuel.
\newblock {\em International Journal of Hydrogen Energy}, 17(6):391--404, 1992.

\bibitem{song2024optimizing}
Meng-Chen Song, Fu-Ying Wu, Yi-Qun Jiang, Xiu-Zhen Wang, Hu~Zhao, Li-Xin Chen, and Liu-Ting Zhang.
\newblock Optimizing feconicrti high-entropy alloy with hydrogen pumping effect to boost de/hydrogenation performance of magnesium hydride.
\newblock {\em Rare Metals}, 43(7):3273--3285, 2024.

\bibitem{preethi2013photocatalytic}
V~Preethi and SJMSISP Kanmani.
\newblock Photocatalytic hydrogen production.
\newblock {\em Materials Science in Semiconductor Processing}, 16(3):561--575, 2013.

\bibitem{gong2022perspective}
Yixuan Gong, Jiasai Yao, Ping Wang, Zhenxing Li, Hongjun Zhou, and Chunming Xu.
\newblock Perspective of hydrogen energy and recent progress in electrocatalytic water splitting.
\newblock {\em Chinese Journal of Chemical Engineering}, 43:282--296, 2022.

\bibitem{balat2009production}
M~Balat.
\newblock Production of hydrogen via biological processes.
\newblock {\em Energy Sources, Part A: Recovery, Utilization, and Environmental Effects}, 31(20):1802--1812, 2009.

\bibitem{li2014investigation}
Wenni Li, Qinghai Li, Rui Chen, Yi~Wu, and Yanguo Zhang.
\newblock Investigation of hydrogen production using wood pellets gasification with steam at high temperature over 800 c to 1435 c.
\newblock {\em International journal of hydrogen energy}, 39(11):5580--5588, 2014.

\bibitem{schlapbach2001hydrogen}
Louis Schlapbach and Andreas Z{\"u}ttel.
\newblock Hydrogen-storage materials for mobile applications.
\newblock {\em nature}, 414(6861):353--358, 2001.

\bibitem{jena2011materials}
Puru Jena.
\newblock Materials for hydrogen storage: past, present, and future.
\newblock {\em The Journal of Physical Chemistry Letters}, 2(3):206--211, 2011.

\bibitem{liu2020trends}
Wenjing Liu, Lu~Sun, Zhaoling Li, Minoru Fujii, Yong Geng, Liang Dong, and Tsuyoshi Fujita.
\newblock Trends and future challenges in hydrogen production and storage research.
\newblock {\em Environmental science and pollution research}, 27:31092--31104, 2020.

\bibitem{garcia2018benchmark}
Paula Garc{\'\i}a-Holley, Benjamin Schweitzer, Timur Islamoglu, Yangyang Liu, Lu~Lin, Stephanie Rodriguez, Mitchell~H Weston, Joseph~T Hupp, Diego~A G{\'o}mez-Gualdr{\'o}n, Taner Yildirim, et~al.
\newblock Benchmark study of hydrogen storage in metal--organic frameworks under temperature and pressure swing conditions.
\newblock {\em ACS Energy Letters}, 3(3):748--754, 2018.

\bibitem{abe2019hydrogen}
John~O Abe, API Popoola, Emmanueal Ajenifuja, and Olawale~M Popoola.
\newblock Hydrogen energy, economy and storage: Review and recommendation.
\newblock {\em International journal of hydrogen energy}, 44(29):15072--15086, 2019.

\bibitem{ZHONG2025148}
Tao Zhong, Tian Xu, Liuting Zhang, Fuying Wu, Yiqun Jiang, and Xuebin Yu.
\newblock Designing multivalent nimn-based layered nanosheets with high specific surface area and abundant active sites for solid-state hydrogen storage in magnesium hydride.
\newblock {\em Journal of Magnesium and Alloys}, 13(1):148--160, 2025.

\bibitem{LI2025109566}
Shuai Li, Liuting Zhang, Fuying Wu, Yiqun Jiang, and Xuebin Yu.
\newblock Efficient catalysis of fenicu-based multi-site alloys on magnesium-hydride for solid-state hydrogen storage.
\newblock {\em Chinese Chemical Letters}, 36(1):109566, 2025.

\bibitem{WANG2025159591}
Li~Wang, Tao Zhong, Fuying Wu, Daifen Chen, Zhengdong Yao, Lixin Chen, and Liuting Zhang.
\newblock Anions intercalated two-dimension high entropy layered metal oxides for enhanced hydrogen storage in magnesium hydride.
\newblock {\em Chemical Engineering Journal}, 505:159591, 2025.

\bibitem{GUEMOU202483}
Samuel Guemou, Liuting Zhang, Shuai Li, Yiqun Jiang, Tao Zhong, Zichuan Lu, Ren Zhou, Fuying Wu, and Qian Li.
\newblock Exceptional catalytic effect of novel rgo-supported ni-nb nanocomposite on the hydrogen storage properties of mgh2.
\newblock {\em Journal of Materials Science \& Technology}, 172:83--93, 2024.

\bibitem{rosi2003hydrogen}
Nathaniel~L Rosi, Juergen Eckert, Mohamed Eddaoudi, David~T Vodak, Jaheon Kim, Michael O'Keeffe, and Omar~M Yaghi.
\newblock Hydrogen storage in microporous metal-organic frameworks.
\newblock {\em Science}, 300(5622):1127--1129, 2003.

\bibitem{langmi2014hydrogen}
Henrietta~W Langmi, Jianwei Ren, Brian North, Mkhulu Mathe, and Dmitri Bessarabov.
\newblock Hydrogen storage in metal-organic frameworks: a review.
\newblock {\em Electrochimica Acta}, 128:368--392, 2014.

\bibitem{martin2018identifying}
A.~Martin-Calvo, J.~J. Guti{\'e}rrez-Sevillano, I.~Matito-Martos, T.~J.~H. Vlugt, and S.~Calero.
\newblock Identifying zeolite topologies for storage and release of hydrogen.
\newblock {\em The Journal of Physical Chemistry C}, 122(23):12485--12493, 2018.

\bibitem{ramirez2007dihydrogen}
AJ~Ramirez-Cuesta, PCH Mitchell, DK~Ross, PA~Georgiev, PA~Anderson, HW~Langmi, and David Book.
\newblock Dihydrogen in cation-substituted zeolites x—an inelastic neutron scattering study.
\newblock {\em Journal of Materials Chemistry}, 17(24):2533--2539, 2007.

\bibitem{kumar2021mxenes}
Pradip Kumar, Shiv Singh, SAR Hashmi, and Ki-Hyun Kim.
\newblock Mxenes: Emerging 2d materials for hydrogen storage.
\newblock {\em Nano Energy}, 85:105989, 2021.

\bibitem{shahzad20202d}
Faisal Shahzad, Aamir Iqbal, Hyerim Kim, and Chong~Min Koo.
\newblock 2d transition metal carbides (mxenes): applications as an electrically conducting material.
\newblock {\em Advanced Materials}, 32(51):2002159, 2020.

\bibitem{https://doi.org/10.1002/adfm.202418230}
Tao Zhong, Tian Xu, Liuting Zhang, Li~Wang, Fuying Wu, and Xuebin Yu.
\newblock Modulation on surface termination groups to optimize the adsorption energy and work function of nb2ctx for enhanced hydrogen storage in magnesium hydride.
\newblock {\em Advanced Functional Materials}, 35(13):2418230, 2025.

\bibitem{WANG20143780}
Guangzhao Wang, Hongkuan Yuan, Anlong Kuang, Wenfeng Hu, Guolin Zhang, and Hong Chen.
\newblock High-capacity hydrogen storage in li-decorated (aln)n (n = 12, 24, 36) nanocages.
\newblock {\em International Journal of Hydrogen Energy}, 39(8):3780--3789, 2014.

\bibitem{sun2006first}
Qiang Sun, Puru Jena, Qian Wang, and Manuel Marquez.
\newblock First-principles study of hydrogen storage on li12c60.
\newblock {\em Journal of the American Chemical Society}, 128(30):9741--9745, 2006.

\bibitem{cheng2001hydrogen}
Hui-Ming Cheng, Quan-Hong Yang, and Chang Liu.
\newblock Hydrogen storage in carbon nanotubes.
\newblock {\em Carbon}, 39(10):1447--1454, 2001.

\bibitem{gopalsamy2014hydrogen}
K~Gopalsamy and V~Subramanian.
\newblock Hydrogen storage capacity of alkali and alkaline earth metal ions doped carbon based materials: A dft study.
\newblock {\em international journal of hydrogen energy}, 39(6):2549--2559, 2014.

\bibitem{jaiswal2022alkali}
Ankita Jaiswal, Rakesh~K Sahoo, Shakti~S Ray, and Sridhar Sahu.
\newblock Alkali metals decorated silicon clusters (sinmn, n= 6, 10; m= li, na) as potential hydrogen storage materials: A dft study.
\newblock {\em International Journal of Hydrogen Energy}, 47(3):1775--1789, 2022.

\bibitem{zhang2024enhanced}
Ningning Zhang, Mingyu Wu, Jiwen Li, Wenting Lv, Jinghua Guo, Yu~Yang, Yujuan Zhang, and Ping Zhang.
\newblock Enhanced hydrogen storage capacity in oli3-decorated holey graphitic carbon nitride monolayer.
\newblock {\em ACS Applied Materials \& Interfaces}, 17(1):1971--1979, 2024.

\bibitem{you2024dft}
A~You, Y~Liu, X~Yue, J~Xiao, J~Du, HZ~Huang, and JG~Song.
\newblock Dft study of the hydrogen adsorption behavior on superalkali nli3m decorated graphyne nanosheet under ambient conditions.
\newblock {\em International Journal of Hydrogen Energy}, 65:515--525, 2024.

\bibitem{naqvi2017hexagonal}
Syeda~Rabab Naqvi, Gollu~Sankar Rao, Wei Luo, Rajeev Ahuja, and Tanveer Hussain.
\newblock Hexagonal boron nitride (h-bn) sheets decorated with oli, ona, and li2f molecules for enhanced energy storage.
\newblock {\em ChemPhysChem}, 18(5):513--518, 2017.

\bibitem{hussain2013hexagonal}
Tanveer Hussain, Abir De~Sarkar, Tae~Won Kang, and Rajeev Ahuja.
\newblock Hexagonal boron nitride sheet decorated by polylithiated species for efficient and reversible hydrogen storage.
\newblock {\em Science of Advanced Materials}, 5(12):1960--1966, 2013.

\bibitem{chen2023reversible}
Xihao Chen, Wenjie Hou, Fuqiang Zhai, Jiang Cheng, Shuang Yuan, Yihan Li, Ning Wang, Liang Zhang, and Jie Ren.
\newblock Reversible hydrogen storage media by g-cn monolayer decorated with nli4: a first-principles study.
\newblock {\em Nanomaterials}, 13(4):647, 2023.

\bibitem{zhang2022hydrogen}
Yafei Zhang and Pingping Liu.
\newblock Hydrogen storage on superalkali nli4 decorated $\beta$12-borophene: A first principles insights.
\newblock {\em International Journal of Hydrogen Energy}, 47(32):14637--14645, 2022.

\bibitem{junior2023irida}
ML~Pereira J{\'u}nior, Wiliam~Ferreira da~Cunha, William~Ferreira Giozza, Rafael~Timoteo de~Sousa~Junior, and LA~Ribeiro Junior.
\newblock Irida-graphene: A new 2d carbon allotrope.
\newblock {\em FlatChem}, 37:100469, 2023.

\bibitem{TAN2024738}
Yongkang Tan, Xiaoma Tao, Yifang Ouyang, and Qing Peng.
\newblock Stable and 7.7 wt
\newblock {\em International Journal of Hydrogen Energy}, 50:738--748, 2024.

\bibitem{ZHANG20241004}
Ya-Fei Zhang and Junxiong Guo.
\newblock Li-decorated 2d irida-graphene as a potential hydrogen storage material: A dispersion-corrected density functional theory calculations.
\newblock {\em International Journal of Hydrogen Energy}, 50:1004--1014, 2024.

\bibitem{YUAN2024114756}
Lihua Yuan, Mengjia Shi, Junyan Su, Daobin Wang, Haimin Zhang, Jijun Gong, and Jinyuan Ma.
\newblock First-principles investigation of irida-graphene decorated with alkali metal for reversible hydrogen storage.
\newblock {\em Computational and Theoretical Chemistry}, 1239:114756, 2024.

\bibitem{DUAN20241}
Zhanjiang Duan, Shunping Shi, Chunyu Yao, Xiaoling Liu, Kai Diao, Dan Lei, and Yiliang Liu.
\newblock Reversible hydrogen storage with na-modified irida-graphene: A density functional theory study.
\newblock {\em International Journal of Hydrogen Energy}, 85:1--11, 2024.

\bibitem{PhysRevLett.77.3865}
John~P. Perdew, Kieron Burke, and Matthias Ernzerhof.
\newblock Generalized gradient approximation made simple.
\newblock {\em Phys. Rev. Lett.}, 77:3865--3868, Oct 1996.

\bibitem{ernzerhof1999assessment}
Matthias Ernzerhof and Gustavo~E Scuseria.
\newblock Assessment of the perdew--burke--ernzerhof exchange-correlation functional.
\newblock {\em The Journal of chemical physics}, 110(11):5029--5036, 1999.

\bibitem{PhysRevB.50.17953}
P.~E. Bl\"ochl.
\newblock Projector augmented-wave method.
\newblock {\em Phys. Rev. B}, 50:17953--17979, Dec 1994.

\bibitem{grimme2010consistent}
Stefan Grimme, Jens Antony, Stephan Ehrlich, and Helge Krieg.
\newblock A consistent and accurate ab initio parametrization of density functional dispersion correction (dft-d) for the 94 elements h-pu.
\newblock {\em The Journal of chemical physics}, 132(15), 2010.

\bibitem{martyna1992nose}
Glenn~J Martyna, Michael~L Klein, and Mark Tuckerman.
\newblock Nos{\'e}--hoover chains: The canonical ensemble via continuous dynamics.
\newblock {\em The Journal of chemical physics}, 97(4):2635--2643, 1992.

\bibitem{bader1985atoms}
Richard~FW Bader.
\newblock Atoms in molecules.
\newblock {\em Accounts of chemical research}, 18(1):9--15, 1985.

\bibitem{BENIWAL2025114947}
Preeti Beniwal and T.J. {Dhilip Kumar}.
\newblock Superalkali oli3 anchored biphenylene for hydrogen storage: Acumen from first-principles study.
\newblock {\em Journal of Energy Storage}, 108:114947, 2025.

\bibitem{elaggoune2024stability}
Warda Elaggoune and Yusuf~Zuntu Abdullahi.
\newblock Stability, electronic and magnetic properties of boronphosphide (bp) graphenylene induced by interstitial atomic doping.
\newblock {\em Journal of Physics and Chemistry of Solids}, 194:112256, 2024.

\bibitem{alhameedi2019metal}
Khidhir Alhameedi, Amir Karton, Dylan Jayatilaka, and Tanveer Hussain.
\newblock Metal functionalized inorganic nano-sheets as promising materials for clean energy storage.
\newblock {\em Applied Surface Science}, 471:887--892, 2019.

\bibitem{hashmi2017ultra}
Arqum Hashmi, M~Umar Farooq, Imran Khan, Jicheol Son, and Jisang Hong.
\newblock Ultra-high capacity hydrogen storage in a li decorated two-dimensional c 2 n layer.
\newblock {\em Journal of Materials Chemistry A}, 5(6):2821--2828, 2017.

\bibitem{kaewmaraya2023ultrahigh}
T~Kaewmaraya, N~Thatsami, P~Tangpakonsab, R~Kinkla, K~Kotmool, C~Menendez, KF~Aguey-Zinsou, and T~Hussain.
\newblock Ultrahigh hydrogen storage using metal-decorated defected biphenylene.
\newblock {\em Applied Surface Science}, 629:157391, 2023.

\bibitem{PhysRevB.6.3637}
Frank~W. Averill.
\newblock Calculation of the cohesive energies and bulk properties of the alkali metals.
\newblock {\em Phys. Rev. B}, 6:3637--3642, Nov 1972.

\bibitem{JIANG2024865}
Minming Jiang, Jiang Xu, Paul Munroe, Zong-Han Xie, and Zhaofeng Chen.
\newblock Light metal decorated graphene-like si2bn monolayers as hydrogen storage media: A dft investigation.
\newblock {\em International Journal of Hydrogen Energy}, 50:865--878, 2024.

\bibitem{beniwal2025superalkali}
Preeti Beniwal and TJ~Dhilip Kumar.
\newblock Superalkali oli3 anchored biphenylene for hydrogen storage: Acumen from first-principles study.
\newblock {\em Journal of Energy Storage}, 108:114947, 2025.

\bibitem{zhang2023potential}
Ningning Zhang, Yongting Shi, Jiwen Li, Yujuan Zhang, Jinghua Guo, Zhenguo Fu, and Ping Zhang.
\newblock Potential applications of oli3-decorated h-bn monosheet for high hydrogen storage.
\newblock {\em Applied Physics Letters}, 123(8), 2023.

\bibitem{LIU2025105802}
Zizhong Liu, Xihao Chen, Yuehong Liao, Longxin Zhang, and José~A.S. Laranjeira.
\newblock First-principles insights of na-decorated b7n5 monolayer for advanced hydrogen storage.
\newblock {\em Surfaces and Interfaces}, 58:105802, 2025.

\bibitem{doi:10.1021/acs.langmuir.4c00779}
Xihao Chen, Liang Zhang, Huaijie Jia, and Peng Gao.
\newblock Computational investigation of a reversible energy storage medium in g-b5n3 decorated by lithium.
\newblock {\em Langmuir}, 40(22):11582--11589, 2024.
\newblock PMID: 38785077.

\bibitem{CHEN2024114445}
Xihao Chen, Che Zhang, Zonghang Liu, Jiwen Li, Donglin Guo, Liang Zhang, Jiang Cheng, Longxin Zhang, Guangzhao Wang, and Peng Gao.
\newblock First-principles investigation of high reversible energy storage medium in li-decorated net-y.
\newblock {\em Journal of Energy Storage}, 103:114445, 2024.

\bibitem{CHEN2024510}
Xihao Chen, Jiwen Li, Longxin Zhang, Ning Wang, Jiang Cheng, Zhenyu Ma, Peng Gao, Guangzhao Wang, Xinyong Cai, Donglin Guo, Jing Xiang, and Liang Zhang.
\newblock Computational evaluation of li-decorated $\alpha$-c3n2 as a room temperature reversible hydrogen storage medium.
\newblock {\em International Journal of Hydrogen Energy}, 62:510--519, 2024.

\bibitem{MARTINS202498}
Nicolas~F. Martins, Ary~S. Maia, José~A.S. Laranjeira, Guilherme~S.L. Fabris, Anderson~R. Albuquerque, and Julio~R. Sambrano.
\newblock Hydrogen storage on the lithium and sodium-decorated inorganic graphenylene.
\newblock {\em International Journal of Hydrogen Energy}, 51:98--107, 2024.

\end{thebibliography}

\end{document}